\documentclass{aa}

\usepackage{scalefnt}
\usepackage[]{aalongtable}
\usepackage{graphicx}
\usepackage{epstopdf}
\usepackage[varg]{txfonts}
\begin{document}

   \title{Origin of the heavy elements in HD 140283.\\ Measurement of europium abundance.\thanks{Based on observations 
within Brazilian time at the Canada-France-Hawaii Telescope (CFHT) 
which is operated by the National Research Council of Canada, the Institut National 
des Sciences de l'Univers of the Centre National de la Recherche Scientique of France, 
and the University of Hawaii; Progr. ID 11AB01.
} }

   \author{
      C. Siqueira Mello Jr.\inst{1,2}
      \and
      B. Barbuy\inst{1}
      \and
      M. Spite\inst{2}
      \and
      F. Spite\inst{2}
          }
\offprints{C. Siqueira Mello Jr. (cesar.mello@usp.br).}

   \institute{
      IAG, Universidade de S\~ao Paulo, Rua do Mat\~ao 1226,
      Cidade Universit\'aria, S\~ao Paulo 05508-900, Brazil
      \and
      GEPI, Observatoire de Paris,  CNRS, UMR 8111, F-92195 Meudon Cedex, France
             }

   \date{Received July 24, 2012; accepted October 4, 2012}

 
  \abstract
{HD 140283 is a nearby (V=7.7) subgiant metal-poor star, extensively analysed in the literature. 
Although many spectra have been obtained for this star, none showed a signal-to-noise (S/N) ratio 
high enough to enable a very accurate derivation of abundances from weak lines.}
{The detection of europium proves that the neutron-capture elements in this star originate in the r-process, and not
in the s-process, as recently claimed in the literature.}
{Based on the OSMARCS 1D LTE atmospheric model and with a consistent approach based on the spectrum synthesis code Turbospectrum, we measured the europium lines at 
4129 {\AA} and 4205 {\AA}, taking into account the hyperfine structure of the transitions. The spectrum, obtained with a long exposure time of seven hours at 
the Canada-France-Hawaii Telescope (CFHT), has a resolving power of 81000 and a S/N ratio of 800 at 4100 {\AA}.}
{We were able to determine the abundance A(Eu)=-2.35$\pm$0.07 dex, compatible with the value predicted for the europium from the r-process. 
The abundance ratio [Eu/Ba]=+0.58$\pm$0.15 dex agrees with the trend observed in metal-poor stars and is also compatible with a strong r-process contribution to 
the origin of the neutron-capture elements in HD 140283.}
   {}


   \maketitle
%

\section{Introduction}

The neutron-capture element abundances in extremely metal-poor (EMP) stars predominantly originate in the r-process, 
since the s-process in metal-poor stars is significant only in later phases of the Galaxy due to the evolutionary timescales of the proposed sites and their elemental composition. 
Indeed, \citet{Francois2007} found that the s-process begins to increase relative to the r-process only when [Fe/H] reaches -2.6 dex 
(or even {\it at higher metallicities} in some recent scenarios). This assumption was first suggested by \citet{Truran1981} from a theoretical point of view, 
using the results by \citet{Spite1978} on the behaviour of s-elements vs. [Fe/H].

Recently, \citet{Gallagher2010} challenged this interpretation by analysing the isotopic 
fractions of barium in the well-studied metal-poor halo subgiant star HD 140283. 
The $^{134}$Ba and $^{136}$Ba isotopes are produced by the s-process only, whereas $^{135}$Ba and $^{137}$Ba are produced by both the 
s- and r-processes. The $^{138}$Ba isotope is the dominant isotope and is produced by the s-process in the classical approach, with a small contribution from the r-process 
in the stellar model \citep{Arlandini1999}.

Using very high resolution and high S/N spectra, the authors found a barium isotopic fraction that indicates a 100$\%$ contribution by the s-process, 
which contradicts Truran's theory since the metallicity of HD 140283, [Fe/H]=-2.50$\pm$0.20 \citep{Aoki2004}, presumably indicates the absence of s-process contribution, 
which is a long-standing problem \citep{Magain1995,Lambert2002,Collet2009}. 

Very recently, \citet{Gallagher2012} examined the barium isotopic fraction in another five metal-poor stars and found that all of them show a high s-process signature, 
but the [Ba/Eu] ratios found in two stars from the sample indicate a large r-process contribution, which led the authors to propose that it is much more likely that the 
1D LTE techniques employed in the barium isotope analysis are inadequate (due the asymmetric formation of lines) than to believe that all stars analysed disagree 
with the theory. 
However, \citet{Collet2009} measured the fractional abundance of odd Ba isotopes using a 3D hydrodynamical model atmosphere of HD 140283 and found a 
contribution of the s-process in this star stronger than the result obtained with the 1D analysis, showing that the problem is not simply due to the 3D-1D correction.

To solve the question of the origin of the neutron-capture elements in HD 140283 independently of the isotopic analysis, in the present paper we derive the precise abundance of the 
r-process element europium in this star, which was extensively analysed in the literature. Although many 
spectra have been obtained for this star, none showed a S/N ratio high enough to derive reliable abundances from weak lines, including the europium abundance. 
This work is organized as follows: Sect. 2 describes the observations and data reduction; Sect. 3 summarises the procedures of abundance determination; 
Sect. 4 discusses the results; Sect. 5 summarises our conclusions.


\section {Observations}

HD 140283 was observed during the programme 11AB01 (PI: B. Barbuy) at Canada-France-Hawaii Telescope (CFHT) with the spectrograph ESPaDOnS in Queued Service Observing (QSO) mode 
to obtain a spectrum in the wavelength range 3700 {\AA} - 10475 {\AA} with a resolution of R=81000. 
The observations were carried out in 2011, on June 12, 14, 15, and 16 and on July 8. 
The total quantity of 23 individual spectra with 20 minutes of exposure each produced the long exposure time of more than seven hours. 
The co-added spectrum was obtained after radial-velocity correction 
and its S/N ratio reaches 800-3400 per pixel. Three spectra were discarded because of their low quality compared with the average. 


\section{Abundance determination}

\subsection{Atmospheric parameters and spectrum synthesis}

\begin{figure}
\centering
\resizebox{70mm}{!}{\includegraphics{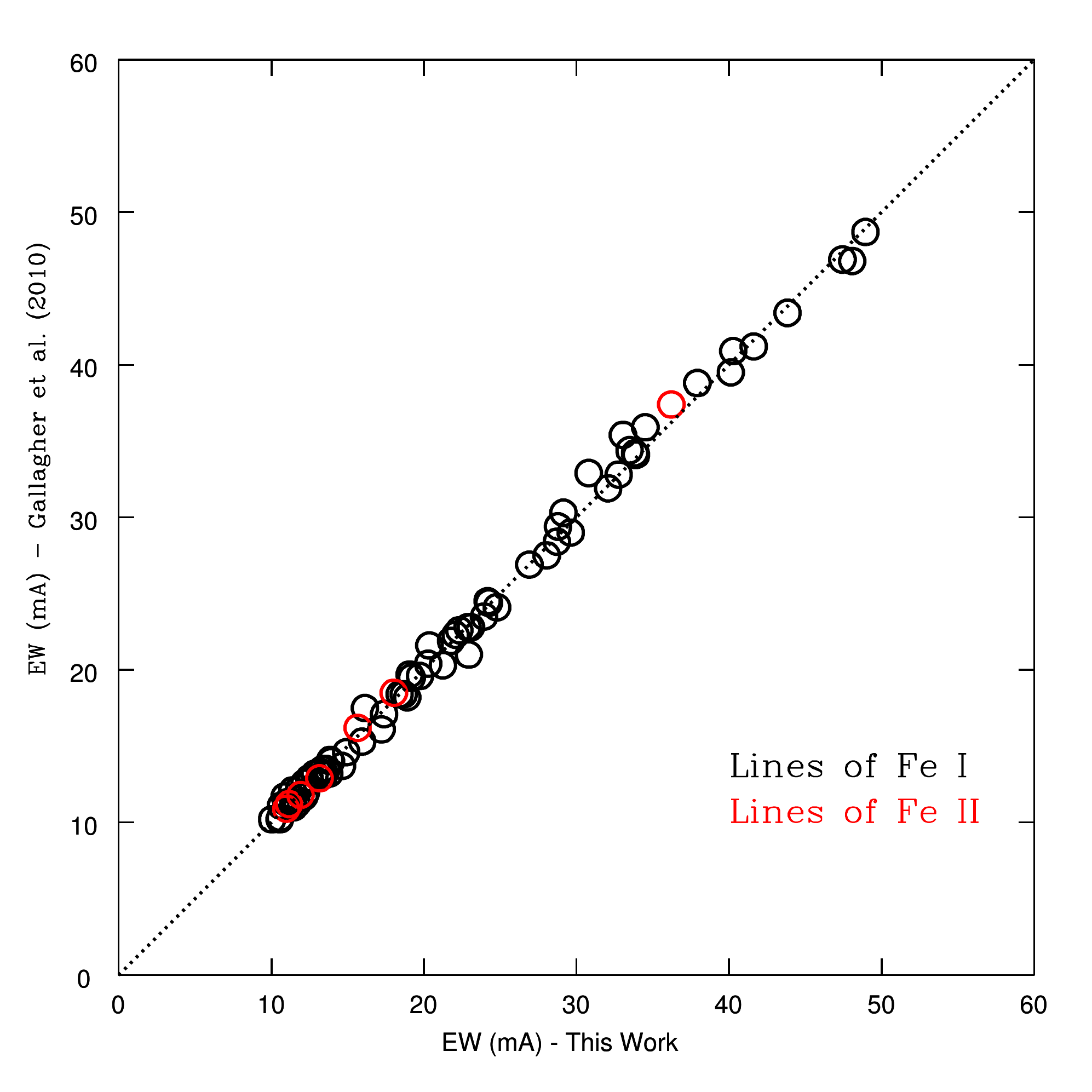}}
\caption{Comparison of the equivalent widths of the Fe I and Fe II lines measured in this work with those from \citet{Gallagher2010}.}
\label{EW}
\end{figure}

The present abundance determination is based on the OSMARCS LTE atmospheric model \citep{Gustafsson2003,Gustafsson2008}, 
which used an updated version of the MARCS program \citep{Gustafsson1975,Plez1992,Asplund1997} to build 1D LTE plane-parallel 
models for cool stars. We used a consistent approach based on the spectrum synthesis code Turbospectrum \citep{Alvarez1998}, which includes a full 
chemical equilibrium and Van der Waals collisional broadening by H, He, and H$_{2}$ following \citet{Anstee1995}, \citet{Barklem1997}, and \citet{Barklem1998}.  

The calculations used the Turbospectrum molecular line lists, 
described in detail in \citet{Alvarez1998}, together with the atomic line lists from the VALD2 
compilation \citep{Kupka1999}. Following \citet{Gallagher2010}, we adopted the stellar parameters 
T$_{\rm eff}$=5750$\pm$100 K, [Fe/H]=-2.5$\pm$0.2 and $v_{t}$=1.4$\pm$0.1 km.s$^{-1}$ from \citet{Aoki2004} and log$g$=3.7$\pm$0.1 [cgs] from \citet{Collet2009}. 
We also adopted the element abundances determined by \citet{Honda2004}. 

To check the reliability of our new spectrum, we determined the equivalent widths (EW) for the same sample of Fe I and Fe II lines used by \citet{Gallagher2010}, 
excluding those lines with contamination by other lines found around of $\pm$0.3 {\AA} from the Fe line centre, as described in table A.1 of the same paper.
We compare the new EW values with the previous ones from \citet{Gallagher2010}, and Fig. \ref{EW} shows the good agreement between them. 
The iron abundance derived was A(Fe)\footnote{
We adopt the notation A(X) = log $\epsilon$(X) = log n(X)/n(H) + 12, with n = number density of atoms.}=+4.92$\pm$0.01 dex, which leads to a metallicity of 
[Fe/H]=-2.60$\pm$0.09, using the solar abundance of iron A(Fe)$_{\odot}$=+7.52$\pm$0.06 dex from \citet{Caffau2011}, also in good agreement with [Fe/H]=-2.59$\pm$0.06 found in 
\citet{Gallagher2010}, who used a code that employs a $\chi^{2}$ test. We present the data for each line in the online material. 
The error in our result is caused by the quadratic sum of the 
uncertainties from the solar abundance and from the iron abundance in the star, which in turn takes into account the observational error and the uncertainties due to atmospheric 
parameters, as discussed in the next section. 

\begin{figure}
\centering
\resizebox{90mm}{!}{\includegraphics{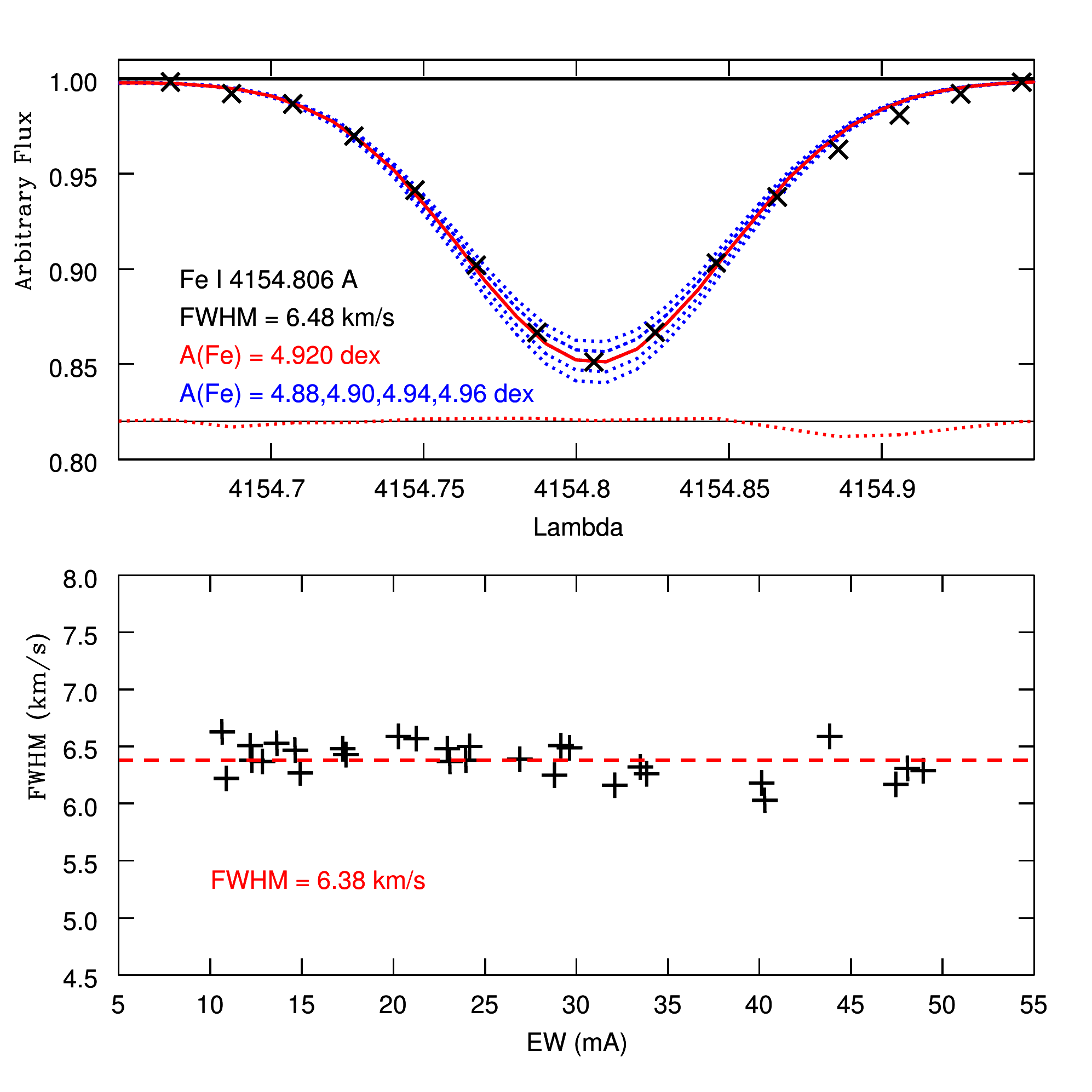}}
\caption{{\bf Upper panel:} fit of the observed Fe I 4154.806 {\AA} line in HD 140283. Crosses: observations. Dotted blue lines: synthetic spectra computed for the
abundances indicated in the figure. Solid red line: synthetic spectrum computed with the best abundance, also indicated in the figure. Dotted red line: 
residual value. {\bf Lower panel:} plot of individual FWHM values for each of the 29 Fe I lines as a function of the equivalent widths. Dashed red line: average value 
FWHM=6.38 km.s$^{-1}$.}
\label{convol}
\end{figure}

Several authors have carefully analysed the broadening parameters in HD 140283, including rotation. 
Recently, \citet{Gallagher2010} determined $v$ sin$i <$ 3.9 km.s$^{-1}$ as a new limit. 
We convolved the synthetic spectrum with a Gaussian profile that 
takes into account the effects of macroturbulence, rotational, 
and instrumental broadening. 
We measured the FWHM of 29 Fe I lines and found an average value of FWHM=6.38 km.s$^{-1}$, which was applied to convolve our synthetic spectra. 
Fig. \ref{convol} shows a typical fit to an iron line (upper panel), and the plot of individual FWHM values for each line as a function of the 
equivalent widths (lower panel). Table \ref{felinesconvol} gives the wavelength and FWHM for each line. 
Note that we are using lines with EW$<$50 m{\AA} to avoid the uncertain computation of broadening effects on the line wings, 
as well as with EW$>$10 m{\AA} to avoid introducing weak lines in the average, which are uncertain due to the S/N ratio.

In the upper panel of Fig. \ref{convol} it is also possible to note the typical asymmetric residual in red wings of the fit on the iron lines, reported by other authors, 
which shows the difficulties in fitting absorption lines with 1D LTE synthetic profiles. 
\citet{Gallagher2012} investigated the NLTE effect in iron lines and reported a correction of 
$\Delta \rm FWHM\sim$-0.1 km.s$^{-1}$ on the convolution parameter, 
but they did not find a better fit to the iron lines, which led the authors to claim the need to include other mechanisms to reproduce the observational data.

\begin{table}
\caption{List of iron lines selected from \citet{Gallagher2010} that were used in this work to set the broadening parameter.}             
\label{felinesconvol}      
\centering                          
\begin{tabular}{c c c c}        
\hline\hline                 
\hbox{$\rm \lambda$({\AA})} & \hbox{FWHM (km.s$^{-1}$)} & \hbox{$\rm \lambda$({\AA})} & \hbox{FWHM (km.s$^{-1}$)} \\    
\hline                        
 4132.899 & 6.27 & 4219.360 & 6.38 \\
 4134.678 & 6.39 & 4222.213 & 6.49 \\
 4136.998 & 6.63 & 4225.454 & 6.37 \\
 4143.415 & 6.26 & 4233.603 & 6.59 \\
 4147.669 & 6.37 & 4238.810 & 6.59 \\
 4154.499 & 6.48 & 4282.403 & 6.17 \\
 4154.806 & 6.48 & 4430.614 & 6.53 \\
 4157.780 & 6.47 & 4442.339 & 6.51 \\
 4175.636 & 6.57 & 4443.194 & 6.51 \\
 4184.892 & 6.43 & 4447.717 & 6.50 \\
 4187.039 & 6.31 & 4461.653 & 6.03 \\
 4191.431 & 6.18 & 4466.552 & 6.25 \\
 4199.095 & 6.29 & 4489.739 & 6.22 \\
 4210.344 & 6.16 & 4494.563 & 6.32 \\
 4217.546 & 6.38 &          &      \\
\hline                                   
\end{tabular}
\end{table}

\subsection{Uncertainties on the derived abundances}

The adopted atmospheric parameters present typical errors: $\Delta$T$_{\rm eff}$=100 K, $\Delta$log $g$=0.1 [cgs], and $\Delta v_{t}$=0.1 km.s$^{-1}$.
We estimated the abundance uncertainties arising from each of these three sources independently. The results are shown in
Table \ref{erro_model} (columns 2 to 4), where the models B, C, and D are compared with the nominal model labelled A.

The quadratic sum of the various sources of uncertainties is not the best way to estimate the total error budget, since the stellar parameters are not independent of each other, 
which adds significant covariance terms in this calculation. To avoid estimating the correlation matrix and the introduction of uncontrollable error sources, 
we created a new atmospheric model with a 100 K lower temperature, 
determining the corresponding surface gravity and microturbulent velocity by the traditional method. Requiring that the iron abundance derived from the Fe I and Fe II lines be 
identical, we determined the respective log$g$ value, and the microturbulent velocity was found requiring that the abundances derived for individual Fe I lines be independent
of the equivalent widths of the lines. The result is a model with T$_{\rm eff}$=5650 K, log $g$=3.3 [cgs], and $v_{t}$=1.2 km.s$^{-1}$, designated as model E in 
Table \ref{erro_model}, and the difference with the nominal model should represent the total error budget arising from the stellar parameters, presented in columns 5 of 
Table \ref{erro_model}.

Observational errors in the cases of vanadium and iron were estimated using the standard deviation of the abundances from the individual lines for each element, 
and must take into account the uncertainties in defining the continuum, fitting the line profiles, and in the oscillator strengths. 
For carbon we used the error found for vanadium as a good representation since this approximation is not expected to lead to a significant difference. 
Finally, the observational error of europium was determined with a specific methodology, as described in section 3.4.

\begin{table}
\caption{Abundance uncertainties due to stellar parameters.}             
\label{erro_model}      
\centering                          
\begin{tabular}{c c c c c}        
\hline\hline                 
\multicolumn{5}{c}{A: $T_{\rm eff}$ = 5750, log $g$ = 3.7 dex, $v_{t}$ = 1.4 km s$^{-1}$} \\
\multicolumn{5}{c}{B: $T_{\rm eff}$ = 5750, log $g$ = 3.6 dex, $v_{t}$ = 1.4 km s$^{-1}$} \\
\multicolumn{5}{c}{C: $T_{\rm eff}$ = 5750, log $g$ = 3.7 dex, $v_{t}$ = 1.3 km s$^{-1}$} \\
\multicolumn{5}{c}{D: $T_{\rm eff}$ = 5650, log $g$ = 3.7 dex, $v_{t}$ = 1.4 km s$^{-1}$} \\
\multicolumn{5}{c}{E: $T_{\rm eff}$ = 5650, log $g$ = 3.3 dex, $v_{t}$ = 1.2 km s$^{-1}$} \\
\hline\hline                 
\hbox{El.} & \hbox{$\Delta_{B-A}$} & \hbox{$\Delta_{C-A}$} & \hbox{$\Delta_{D-A}$} & \hbox{$\Delta_{E-A}$} \\
\hline                        
\hbox{[Fe/H]}  & ~~~~~~0.00  & ~~+0.01 & ~~~-0.08 & ~~~-0.07~~ \\
\hbox{[C/Fe]}  & ~~+0.03 & ~~-0.02 & ~~~-0.11 & ~~~+0.01~~ \\
\hbox{[V/Fe]}  & ~~-0.01 & ~~-0.02 & ~~~+0.04 & ~~~-0.09~~ \\
\hbox{[Ba/Fe]} & ~~-0.04 & ~~~~~~0.00 & ~~~+0.01 & ~~~-0.09~~ \\
\hbox{[Eu/Fe]} & ~~-0.07 & ~~-0.03 & ~~~+0.01 & ~~~-0.05~~ \\
\hline                                   
\end{tabular}
\end{table}

\subsection{Verification of blends}

The europium lines used in the present work are weak and surrounded by regions with CH bands. 
To guarantee the reliability of our results, we first checked the quality of the CH bands in our synthetic spectrum. 

\begin{figure}
\centering
\resizebox{70mm}{!}{\includegraphics{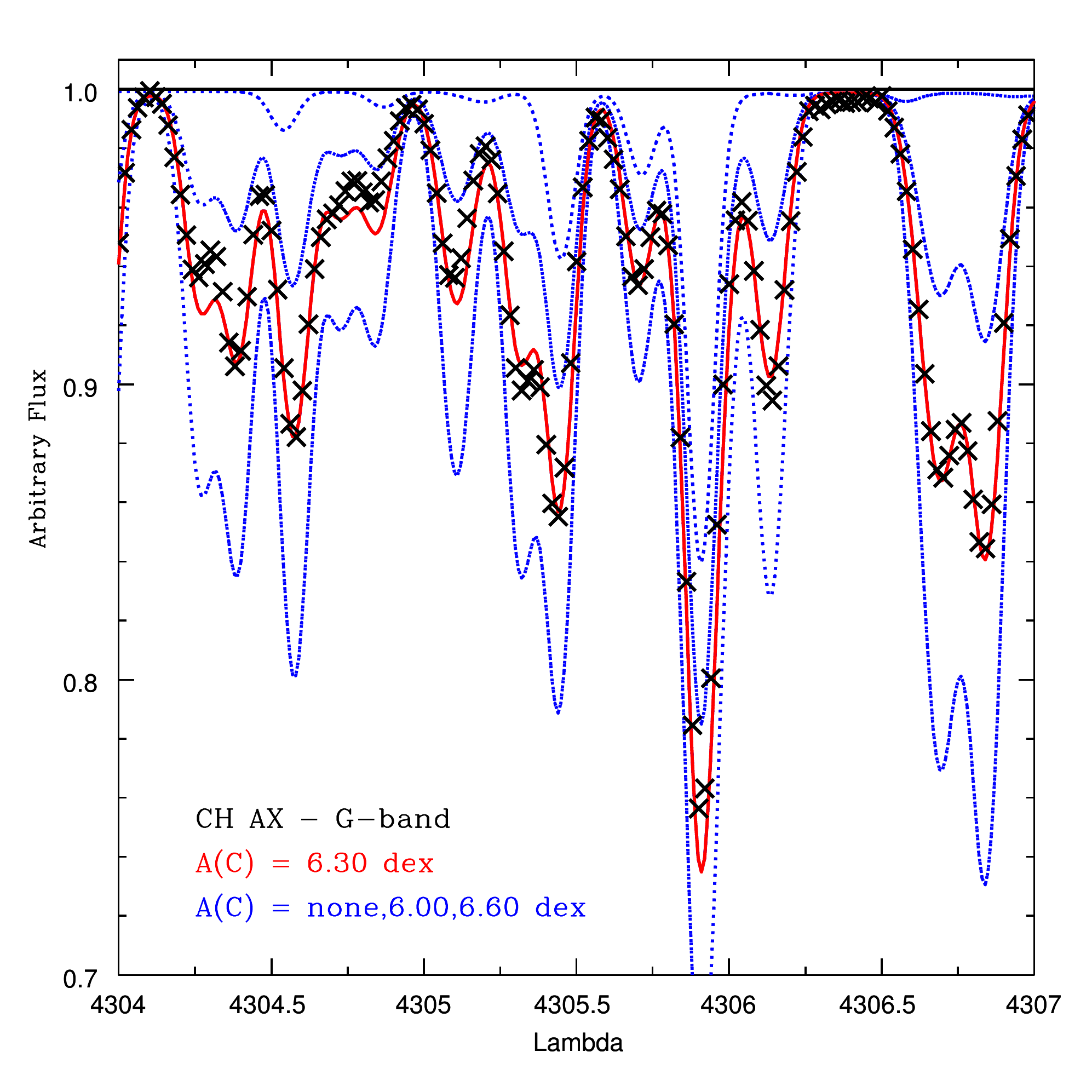}}
\caption{Fit of the observed CH lines (G band) in HD 140283. 
Crosses: observed spectrum. 
Dotted lines: synthetic spectra computed for the abundances 
indicated in the figure. 
Solid line: synthetic spectrum computed with the abundance giving the best fit, also indicated in the figure.}
\label{gband}
\end{figure}

Fig. \ref{gband} shows the fit of the observed lines of CH AX electronic transition band (G band) in HD 140283. 
One can see that the spectrum is well-described by the model using the carbon abundance A(C)=+6.30 dex adopted from \citet{Honda2004}. 
Indeed, several CH bands were analysed and all of them presented a good fit, therefore we assume that they are properly taken into account and have no 
influence on the derivation of the europium abundance. 

\begin{figure}
\centering
\resizebox{90mm}{!}{\includegraphics{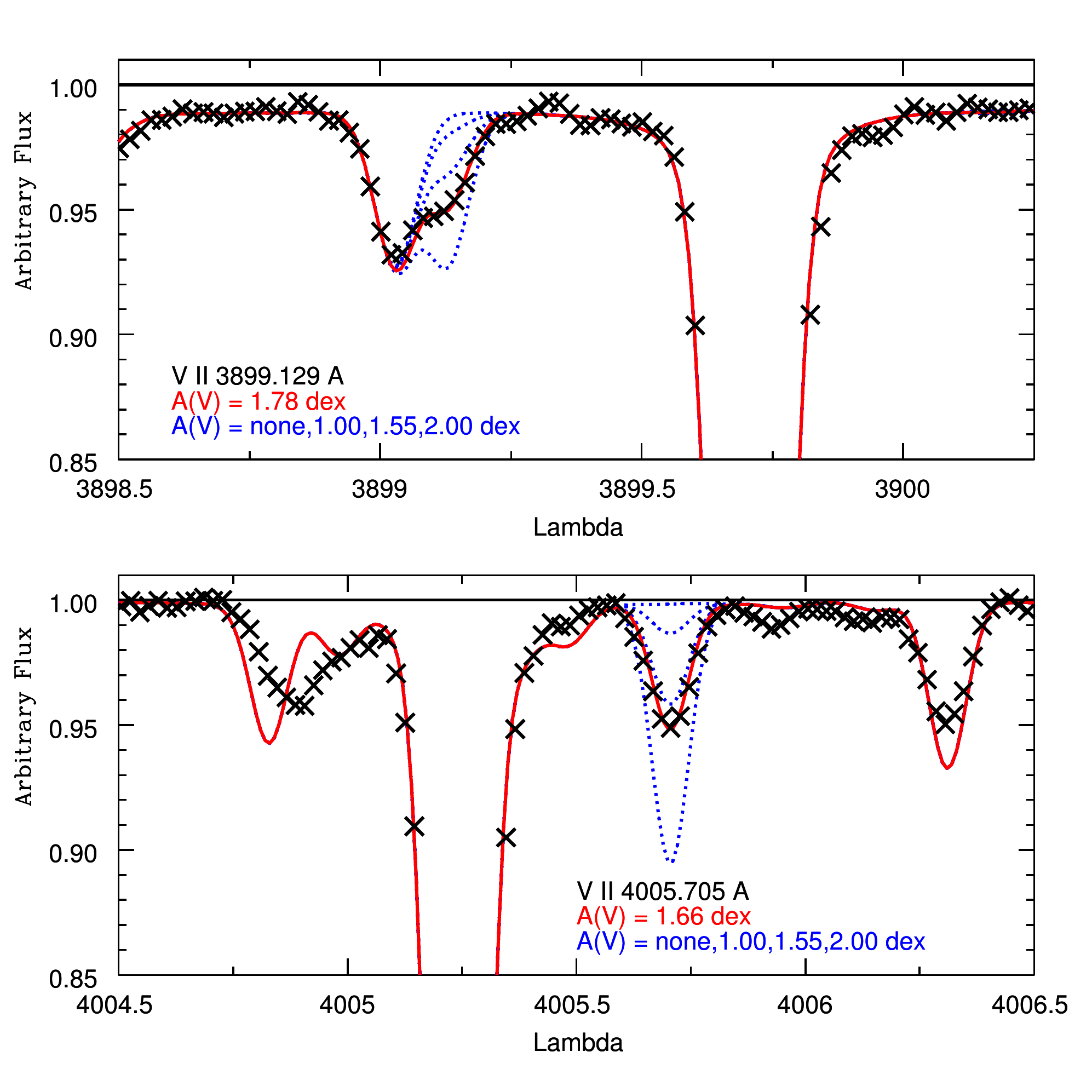}}
\caption{Fit of the observed V II 3899.129 {\AA} and V II 4005.705 {\AA} lines in HD 140283. Crosses: observed spectrum. 
Dotted lines: synthetic spectra computed for 
the abundances indicated in the figure. Solid line: synthetic spectrum computed with the abundance giving the best fit, also indicated in the figure.}
\label{v}
\end{figure}

Another important element to be checked is vanadium. 
Only two europium lines are strong enough to be used in this star, and the line Eu II 4205.05 {\AA} is blended with V II 4205.08 {\AA}. 
Therefore, a correct determination of the abundance of this element is fundamentally important for the europium analysis. 
Table \ref{vlines} summarises our line list and the individual abundances of the V I and V II lines, and Fig. \ref{v} shows examples of 
fits of computed to observed lines. 
From the five V I transitions we obtain A(V)=+1.35$\pm$0.10 dex, while the seven 
V II lines give A(V)=+1.72$\pm$0.10 dex, higher than the result from non-ionized state. Finally, using the complete set of lines we obtain A(V)=+1.56$\pm$0.11 
dex as the final abundance, in excellent agreement with A(V)=+1.55 from \citet{Honda2004}, derived in their analysis from three lines 
(marked with an asterisk in Table \ref{vlines}).

\begin{table}
\caption{List of vanadium lines with the individual abundances. $^{*}$: lines used by \citet{Honda2004}. 
$^{**}$: line reported as not detected by \citet{Honda2004}.}             
\label{vlines}      
\centering                          
\begin{tabular}{c c c c c}        
\hline\hline                 
\hbox{$\rm \lambda$({\AA})} & \hbox{$\chi_{\rm ex}$ (eV)} & \hbox{log gf} & \hbox{A(V)} &  \hbox{[V/Fe]}\\
\hline                        
\multicolumn{5}{c} {\hbox{V~I}}\\
3855.841 &  0.069 & 0.013 & +1.28 & -0.11 \\
4379.230$^{*}$ &  0.301 & 0.580 & +1.35 & -0.04 \\
4384.712 &  0.287 & 0.510 & +1.27 & -0.12 \\
4389.976$^{**}$ &  0.275 & 0.200 & +1.45 & +0.06 \\
4408.193 &  0.275 & 0.020 & +1.40 & +0.01 \\
\multicolumn{5}{c} {\hbox{V~II}}\\
3899.129 & 1.805 & -0.784 & +1.78 & +0.39 \\
3916.411 & 1.428 & -1.053 & +1.83 & +0.44 \\
3951.960$^{*}$ & 1.476 & -0.784 & +1.60 & +0.21 \\
3997.117 & 1.476 & -1.230 & +1.65 & +0.26 \\
4002.936 & 1.428 & -1.447 & +1.75 & +0.36 \\
4005.705$^{*}$ & 1.817 & -0.522 & +1.66 & +0.27 \\
4023.378 & 1.805 & -0.689 & +1.75 & +0.36 \\
\hline                                   
\end{tabular}
\end{table}

\subsection{One-dimensional LTE abundance of europium}

The europium abundance indicators in HD 140283 are the two Eu II resonance lines 4129.70 {\AA} and 4205.05 {\AA}. We computed synthetic spectra as described in section 3.1, 
using the hyperfine structure for the europium transitions from Kurucz\footnote{http://kurucz.harvard.edu/atoms/6301/}. 

\begin{figure}
\centering
\resizebox{93mm}{!}{\includegraphics{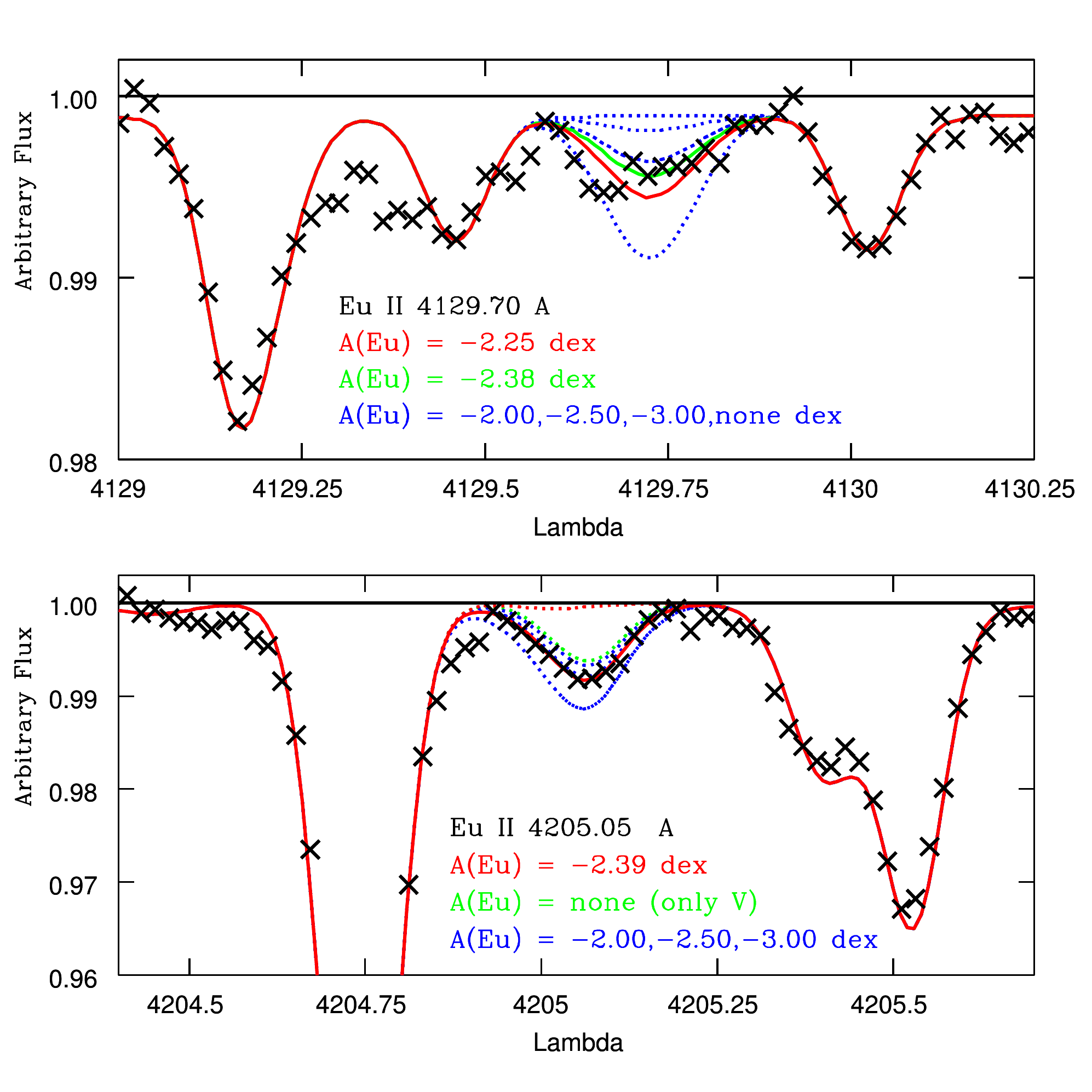}}
\caption{Fit of the observed Eu II 4129.70 {\AA} and Eu II 4205.05 lines in HD 140283. 
Crosses: observed spectrum. Dotted lines: synthetic spectra computed for the abundances indicated in the figure (blue), only with the contribution of V abundance (green), and 
without Eu and V (red). Solid line: synthetic spectrum computed with the best suited Eu abundance (red) and with a minimum limit (green), also indicated in the figure.}
\label{eu}
\end{figure}

The line Eu II 4129.70 {\AA} does not present identified blends, but the position of the continuum is defined by 
the wing of the H$\delta$ hydrogen line at 4101.71 {\AA}. 
In addition, one can see in the upper panel of Fig. \ref{eu} that there is a small structure in the blue wing of this line that could be interpreted as an unidentified 
blend in this region. We checked the spectra of HD 140283 obtained with other instruments, but the resolutions and the S/N ratios of the available data are not sufficient 
to allow us to resolve this absorption profile. 
From the list of known lines, the most compatible with the wavelength of the asymmetry is Ti I 4129.643 {\AA}, but it is necessary to artificially increase 
its $gf$-value by a factor of almost 100 to fit the profile, and in this case we found A(Eu)=-2.43 dex. 

More optimistic, noise is another interpretation of the asymmetric structure and in this case a higher europium abundance is required to explain the profile, 
leading us to find A(Eu)=-2.25 dex, represented by the solid red line in Fig. \ref{eu} (upper panel). 
However, it is also possible to interpret the region as an observational problem and then the solid green line in Fig. \ref{eu} is the best fit to the data, 
compatible with A(Eu)=-2.38 dex. We decided to assume these variations inside the observational error and adopted the average of the three results A(Eu)=-2.35$\pm$0.07 dex as 
the final abundance. Using the solar abundace of europium A(Eu)$_{\odot}$=+0.52$\pm$0.03 dex from \citet{Caffau2011}, we have [Eu/Fe]=-0.27$\pm$0.12 dex.

On the other hand, as described in section 3.3, the V II 4205.08 {\AA} line is the dominant component in Eu II 4205.05 {\AA}, which is characterized by the atomic data 
log$gf$=-1.30 and $\chi_{\rm ex}$=+2.04 eV. 
We initially computed the synthetic spectrum of this region taking only the vanadium contribution A(V)=+1.56 dex into account, 
to check the influence of this element in the line profile. 
The dotted green line in Fig. \ref{eu} (lower panel) shows the result, from which one can see that the contribution of europium abundance is important to explain the observed 
intensity of the line. The dotted red line in the same figure is the calculation without europium and vanadium. 
We found A(Eu)=-2.39$\pm$0.08 dex for the europium abundance, or [Eu/Fe]=-0.31$\pm$0.12 dex, in good agreement with the previous line.

\begin{figure}
\centering
\resizebox{93mm}{!}{\includegraphics{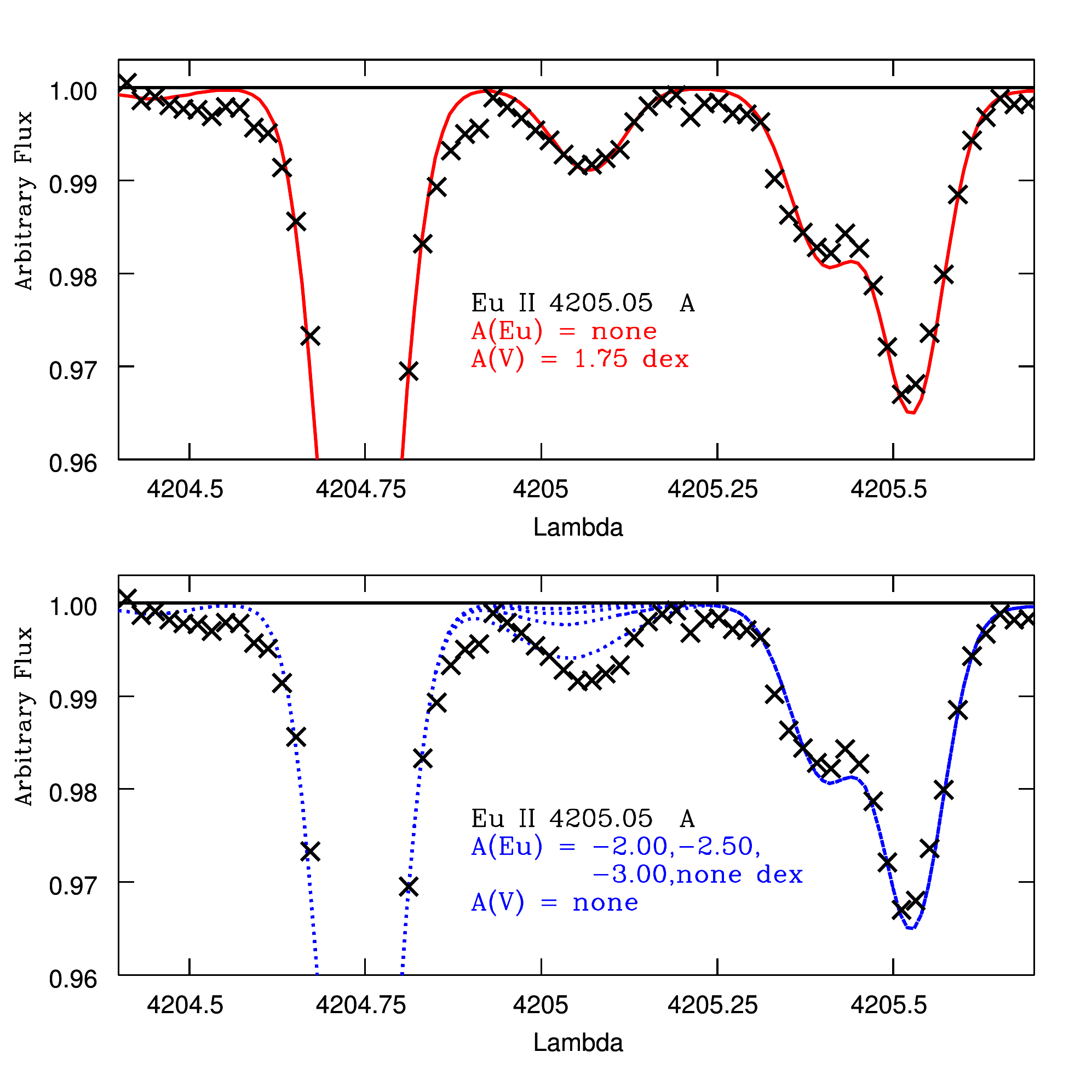}}
\caption{Fit of the observed Eu II 4205.05 {\AA} line in HD 140283. 
Crosses: observed spectrum. Solid red line: synthetic spectrum computed with A(V II)=1.72 dex. Dotted blue lines: synthetic spectra computed only with the contribution of 
Eu, with the abundances indicated in the figure.}
\label{eu2semV}
\end{figure}

Since the vanadium line blending Eu II 4205.05 {\AA} is a transition from the ionized state, we also checked the synthetic spectrum of this profile using the 
higher vanadium abundance A(V)=+1.72 dex obtained with the V II lines. As shown in the upper panel of Fig. \ref{eu2semV}, the higher value of vanadium makes the profile 
compatible with the absence of europium, but one can see an asymmetry in the blue wing of the observed spectrum. 
Although there may be some contamination from an unknown line, looking only at the contribution of the europium line in this region 
(lower panel in Fig. \ref{eu2semV}), it seems reasonable to assume that this component is necessary to reproduce the asymmetric profile. 

In addition, the vanadium abundance from V II lines has a standard deviation $\sigma$=0.08 dex, which means that it is possible to find lower 
abundances from a given line inside a certain probability, for instance A(V)=+1.60 dex (as the line V II 3951.960 {\AA} in Table \ref{vlines}) with almost 7\% of probability, 
assuming a Gaussian distribution. 
Therefore we assume the result A(Eu)=-2.39 dex from the Eu II 4205.05 {\AA} line as an upper limit for the Eu abundance, which is consistent with the more robust detection 
from the Eu II 4129.70 {\AA} line.

The new spectrum used in the present work also allows us to check a third line of europium: Eu II 3819.67 {\AA}. 
As weak as the other Eu lines used, the profile is blended with other faint lines and is in the blue wing of the strong iron line Fe I 3820.42 {\AA}, 
which, added to the lower S/N ratio in the observed spectrum, does not allow to use this line as an indicator of Eu abundance; 
but, the synthetic spectrum calculated with the Eu value from the Eu II 4129.70 {\AA} line seems in general agreement with the observed data in this region.

Europium presents two stable isotopes, $^{151}$Eu and $^{153}$Eu, and the isotopic ratio of 0.5:0.5 for Eu 151:153 was chosen for the calculation described above. 
However, the solar system isotopic ratio of Eu 151:153 is 0.48:0.52, as reported by \citet{Arlandini1999}. 
We decided to check if our europium lines are sufficiently sensitive to detect the variation of these parameters and carried out the analysis of abundance also with 
solar system isotopic ratio, which gave us exactly the same result. Therefore we conclude that the exact value of this ratio is not a relevant factor in this case.

\subsubsection{Comparison with previous results}

We report previous values of the europium abundance in HD 140283. Table \ref{resumo} summarises a non-exhaustive list of literature results, 
discussed in more detail below.

In a study of the composition of neutron-capture elements in 20 very metal-poor halo stars, 
\citet{Gilroy1988} used spectra with a resolution R$\sim$35,000 and S/N value of at least 100 to derive an abundance of 
[Eu/Fe]=+0.09$\pm$0.20 dex. The error presented is the typical value for the heavy elements reported by the authors. 
The model atmosphere grids employed were interpolated in those by \citet{Gustafsson1975} and \citet{Bell1976}. 
To take into account the hyperfine splitting structures of the Eu lines, the authors adopted the line data summarised by 
\citet{Sneden1983} with the line analysis code of \citet{Sneden1973}.

\citet{Magain1989} also used spectra with S/N$\approx$100 and R$\sim$15,000 (FWHM) to find [Eu/Fe]=+0.21$\pm$0.16 dex for the europium abundance in HD 140283. 
This value is the result of forcing the computed equivalent widths to agree with the measured ones, and the models were computed with a version of the MARCS 
code \citep{Gustafsson1975}. Given that an uncertainty on the europium abundance is not reported by \citet{Magain1989}, 
we estimate the error by applying the influence of the atmospheric parameters on the ratio [Eu/Fe], together with the observation uncertainties 
of iron lines, as a lower limit. 
On the other hand, \citet{Gratton1994} used spectra with higher resolutions and higher S/N value, around 50000 and 150 respectively, 
to report the abundance [Eu/Fe]=+0.09$\pm$0.17 dex. We estimated the error in the same way as above. 
The atmospheric model used was the same as in \citet{Gilroy1988}. The authors took hyperfine splitting structures into account by adopting data from \citet{Steffen1985}.

Recently, \citet{Gallagher2010} used a very high resolution (R$\sim$95,000) and very high S/N (S/N=870-1110) spectrum to determine an upper 
limit on the europium abundance, [Eu/Fe]$<$-0.21 dex. The synthetic spectra were produced using KURUCZ06 model atmospheres in conjunction with the 1D LTE code ATLAS, 
and the Eu line lists were constructed using hyperfine splitting information from \citet{Krebs1960} and \citet{Becker1993}. In addition, the isotopic ratio 
0.5:0.5 for Eu 151:153 was adopted. Gallagher and coworker's result questioned the previous values for the europium abundance in this star, 
as well as the origin of the heavy elements found in its atmosphere, justifying the need for a new analysis of europium in this star.

The new abundance of europium presented in this work agrees with the limit set by \citet{Gallagher2010}, showing that high-quality spectra are crucial 
for determining reliable abundances from weak lines. The higher values obtained in previous works for the Eu abundance in this star are likely caused by the difficulties 
in defining the profile of the line due to the noise in the region.

\begin{table}
\caption{Atmospheric parameters and europium abundances from previous works on HD 140283.}             
\label{resumo}      
\scalefont{0.85}
\centering                          
\begin{tabular}{c c c c c c}        
\hline\hline                 
\hbox{Reference} & \hbox{T$_{\rm eff}$ ({\rm K})} & \hbox{log$g$} & \hbox{$v_{t}$ (km.s$^{-1}$)} & \hbox{[Fe/H]} & \hbox{[Eu/Fe]}  \\
\hline                        
\hbox{G88}  & 5650 & 3.3  & 1.5 & -2.3  & +0.09$\pm$0.20 \\
\hbox{M89}  & 5640 & 3.1  & 1.5 & -2.75 & +0.21$\pm$0.16 \\
\hbox{GS94} & 5690 & 3.58 & 0.6 & -2.50 & +0.09$\pm$0.17 \\
\hbox{GA10} & 5750 & 3.70 & 1.4 & -2.59 & <-0.21 \\
\hbox{TW} & 5750 & 3.70 & 1.4 & -2.60 & -0.27$\pm$0.12 \\
\hline                                   
\end{tabular}
\tablebib{
(G88)~\citet{Gilroy1988}; (M89)~\citet{Magain1989}; (GS94)~\citet{Gratton1994}; (GA10)~\citet{Gallagher2010}; (TW)~This work.}
\end{table}

\subsection{Barium abundance}

\citet{Gallagher2010} reported a barium abundance of A(Ba)=-1.29$\pm$0.08 dex from Ba II 4554 {\AA} and 4934 {\AA} lines, or [Ba/H]=-3.46$\pm$0.11 dex 
using the solar abundance of barium A(Ba)$_{\odot}$=+2.17$\pm$0.07 dex from \citet{Lodders2009}. 
Barium is an important element to evaluate the origin of the heavy elements in metal-poor stars, therefore we also derived its abundance using our new spectrum. 
We employed the same lines as \citet{Gallagher2010} to obtain A(Ba)=-1.27$\pm$0.11 dex ([Ba/H]=-3.45$\pm$0.13 dex), in good agreement with their value.

%

\section{Discussion}

We presented a genuine detection of europium in HD 140283 based on a spectrum with the same very high quality as the one used by \citet{Gallagher2010} in their analysis. 
From Eu II 4129.70 {\AA} we adopted A(Eu)=-2.35$\pm$0.07 dex as the final abundance, which is consistent with the upper limit A(Eu)=-2.39 dex estimated from the 
Eu II 4205.05 {\AA} line. Using the solar abundance of europium A(Eu)$_{\odot}$=+0.52$\pm$0.03 dex \citep{Caffau2011}, we obtained [Eu/Fe]=-0.27$\pm$0.12 dex. 
Considering the barium abundance derived by \citet{Gallagher2010} as described in the last section, if Ba is produced by the s-process only, 
then there should be no europium, 
whereas if it is produced exclusively by the r-process, then according to \citet{Simmerer2004} the predicted abundance would be A(Eu)=-2.23 dex. 
Clearly, our result is compatible with this latter r-process case.

It is important to note that \citet{Mashonkina2012} presented NLTE abundance corrections for the Eu II 4129 {\AA} line in cool stars. 
Using a model with T$_{\rm eff}$=5780 K and log $g$=4.4 [cgs], they found $\Delta_{NLTE}$(Eu)=+0.03 dex in stars with solar metallicity. 
On the other hand, the correction of $\Delta_{NLTE}$(Eu)=+0.07 dex was found with a model for [Fe/H]=-3.00, but with T$_{\rm eff}$=5000 K and log $g$=1.5. 
Therefore the authors showed that the NLTE corrections are small for this element and this type of stars, and the general trend is to increase the abundance, which in turn 
leads to an even better agreement between the observational result and the expectation from an r-process origin.

\begin{figure}
\centering
\resizebox{85mm}{!}{\includegraphics{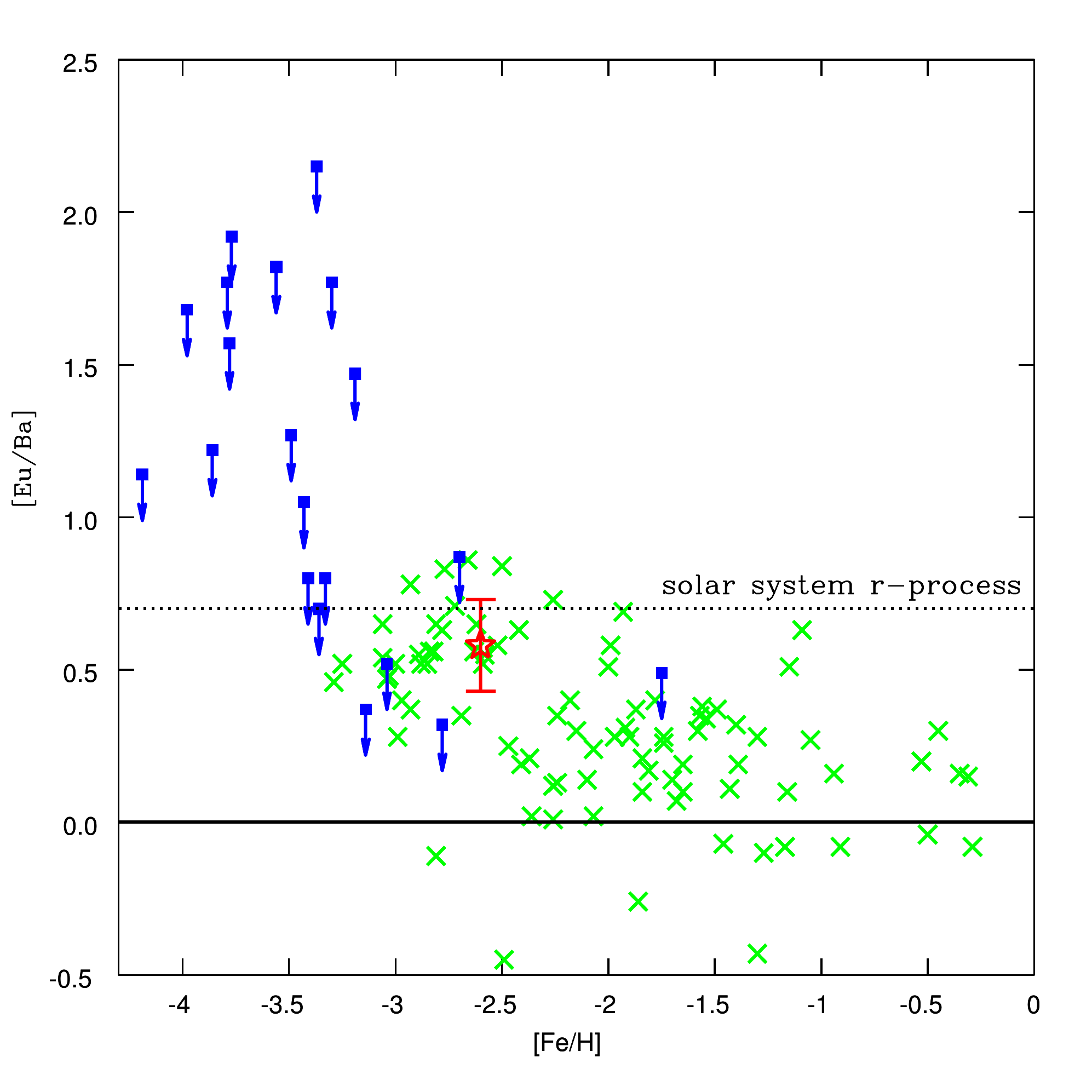}}
\caption{[Eu/Ba] as a function of [Fe/H]. HD 140283 is represented by the red star. 
Green crosses represent data from the literature (see text) for metal-poor stars; 
blue squares are the stars with only an upper limit of europium determined. 
The dotted line indicates the solar system r-process abundance ratio \citep{Simmerer2004}, 
and the solid line is the solar system value.}
\label{compara}
\end{figure}

With the abundances of europium and barium determined in this work we calculated the ratio [Eu/Ba]=+0.58$\pm$0.15 dex in HD 140283. 
The barium abundance obtained by \citet{Gallagher2010} and our new abundance of europium lead to the ratio [Eu/Ba]=+0.60$\pm$0.13 dex, 
showing the consistency of the results, as well as with the upper limit [Eu/Ba]=+0.66 dex set by \citet{Gallagher2010}. 
This ratio is interesting because it allows us to evaluate the relative contribution of the r-process and s-process in the production of elements beyond the iron peak. 
Fig. \ref{compara} compares this abundance ratio in HD 140283 (represented as the red star) with data of other stars selected from previous papers, as 
a function of metallicity \citep{Gilroy1988,Ryan1991,Gratton1994,McWilliam1995,Burris2000,Norris2001,Honda2004,Francois2007}.

The dotted line in Fig. \ref{compara} represents the value [Eu/Ba]$_{r}$=+0.698 dex, associated with a pure r-process contribution according to \citet{Simmerer2004}, whereas 
[Eu/Ba]$_{s}$=-1.621 dex if the abundance pattern is due to the pure s-process. The solid line presents the abundance ratio in the solar system. 
One can see that the ratio found in HD 140283 agrees with the trend observed for metal-poor stars and is compatible 
with a strong r-process contribution and a small s-process component. 
Accordingly, we conclude that the origin of the neutron-capture elements in this star is primarily the rapid neutron-capture process.

Assuming that europium is produced by the r-process only \citep[97$\%$ in the solar system composition according to][]{Simmerer2004}, 
it is possible to estimate the fraction of barium in this star that was produced by the s-process. 
While the values [Eu/Ba]$_{r}$=+0.698 dex and A(Eu)=-2.35 dex lead to A(Ba)=-1.39 dex as the abundance of barium that we expect to be produced exclusively by the r-process, 
the Ba value measured A(Ba)=-1.27 dex means that 11.6$\%$ of this element must be produced by the slow process. 
In addition, if one uses the error obtained for the ratio [Eu/Ba] in this work, the lower value [Eu/Ba]=+0.425 dex leads us to a conservative upper limit of 23.9$\%$ 
for the contribution of the s-process in the barium abundance, since we are ignoring the error associated with the abundance of europium. 
These results may partially explain the result of the Ba isotope ratios obtained by \citet{Gallagher2010,Gallagher2012} and other authors.

Note that our result is within the limit set by \citet{Gallagher2010}, but we present a clearly defined europium abundance together with a careful error analysis, 
which allow one to carry out a valid comparison with predicted values from models and arrive at a different conclusion than \citet{Gallagher2010,Gallagher2012}.

Indeed, from the current picture, it is not very clear when the s-process begins to be important in the evolution of the early Galaxy. 
In Fig \ref{compara} the blue squares represent the stars with only an upper limit for the europium abundance, showing that the efforts to determine the abundance of 
this element in the extremely metal-poor stars are important in the attempt to understand the chemical evolution of the neutron-capture elements in the Galaxy.

%

\section{Conclusions}

With a very high resolution and a very high S/N spectrum we were able to determine a genuine europium abundance in HD 140283. 
From the Eu II 4129.70 {\AA} we found A(Eu)=-2.35$\pm$0.07 dex, in agreement with the predicted strong r-process contribution for the origin of the heavy elements. 

Together with the previous barium abundance \citep{Gallagher2010,Gallagher2012}, confirmed in the present work, 
we obtained the abundance ratio [Eu/Ba]=+0.58$\pm$0.15 dex in this star, 
in agreement with the trend observed in metal-poor stars, and compatible with an r-process origin of the neutron-capture elements in HD 140283.

\begin{acknowledgements}
We thank the referee for his/her useful comments. 
CS and BB acknowledge grants from CAPES, CNPq and FAPESP. 
MS and FS acknowledge the support of CNRS (PNCG and PNPS).
\end{acknowledgements}

\Online

\begin{appendix}
\section{Iron lines}
 
\longtab{1}{
\begin{longtable}{cccccccc}
\caption{\label{felines} List of iron lines selected from \citet{Gallagher2010} that were used in this work.}\\
\hline\hline
\hbox{Spec.} & \hbox{$\rm \lambda$({\AA})} & \hbox{$\chi_{\rm ex}$ (eV)} & \hbox{log gf} & \hbox{EW$_{Gall}$(m{\AA})} 
&  \hbox{EW$_{New}$(m{\AA})} &  \hbox{A(Fe)} &  \hbox{[Fe/H]}\\
\hline
\endfirsthead
\caption{continued.}\\
\hline\hline
\hbox{Spec.} & \hbox{$\rm \lambda$({\AA})} & \hbox{$\chi_{\rm ex}$ (eV)} & \hbox{log gf} & \hbox{EW$_{Gall}$(m{\AA})} 
&  \hbox{EW$_{New}$(m{\AA})} &  \hbox{A(Fe)} &  \hbox{[Fe/H]}\\
\hline
\endhead
\hline
\endfoot
\hbox{Fe~I} & 4132.899 & 2.850 & -1.006 & 14.60 & 14.91 & 4.97 & -2.55 \\
\hbox{Fe~I} & 4134.678 & 2.830 & -0.649 & 26.90 & 26.91 & 4.94 & -2.58 \\
\hbox{Fe~I} & 4136.998 & 3.410 & -0.453 & 11.10 & 10.64 & 4.80 & -2.72 \\
\hbox{Fe~I} & 4143.415 & 3.050 & -0.204 & 34.20 & 33.84 & 4.86 & -2.67 \\
\hbox{Fe~I} & 4147.669 & 1.490 & -2.104 & 22.80 & 23.08 & 4.96 & -2.56 \\
\hbox{Fe~I} & 4153.900 & 3.400 & -0.321 & 18.20 & 18.91 & 4.94 & -2.58 \\
\hbox{Fe~I} & 4154.499 & 2.830 & -0.688 & 21.00 & 22.94 & 4.87 & -2.65 \\
\hbox{Fe~I} & 4154.806 & 3.370 & -0.400 & 16.10 & 17.23 & 4.92 & -2.60 \\
\hbox{Fe~I} & 4157.780 & 3.420 & -0.403 & 13.70 & 14.63 & 4.91 & -2.61 \\
\hbox{Fe~I} & 4175.636 & 2.850 & -0.827 & 20.30 & 21.25 & 4.98 & -2.54 \\
\hbox{Fe~I} & 4184.892 & 2.830 & -0.869 & 17.10 & 17.40 & 4.90 & -2.62 \\
\hbox{Fe~I} & 4187.039 & 2.450 & -0.548 & 46.80 & 48.08 & 4.89 & -2.63 \\
\hbox{Fe~I} & 4191.431 & 2.470 & -0.666 & 39.50 & 40.12 & 4.87 & -2.66 \\
\hbox{Fe~I} & 4199.095 & 3.050 &  0.155 & 48.70 & 48.94 & 4.84 & -2.68 \\
\hbox{Fe~I} & 4210.344 & 2.480 & -0.928 & 31.90 & 32.08 & 4.97 & -2.55 \\
\hbox{Fe~I} & 4217.546 & 3.430 & -0.484 & 12.00 & 12.28 & 4.90 & -2.62 \\
\hbox{Fe~I} & 4219.360 & 3.570 &  0.000 & 23.50 & 23.95 & 4.94 & -2.58 \\
\hbox{Fe~I} & 4222.213 & 2.450 & -0.967 & 29.00 & 29.63 & 4.92 & -2.60 \\
\hbox{Fe~I} & 4225.454 & 3.420 & -0.510 & 13.20 & 12.84 & 4.94 & -2.58 \\
\hbox{Fe~I} & 4233.603 & 2.480 & -0.604 & 43.40 & 43.84 & 4.89 & -2.63 \\
\hbox{Fe~I} & 4238.810 & 3.400 & -0.233 & 20.40 & 20.28 & 4.89 & -2.63 \\
\hbox{Fe~I} & 4282.403 & 2.180 & -0.779 & 46.90 & 47.44 & 4.87 & -2.65 \\
\hbox{Fe~I} & 4430.614 & 2.220 & -1.659 & 13.50 & 13.63 & 4.93 & -2.59 \\
\hbox{Fe~I} & 4442.339 & 2.200 & -1.255 & 30.30 & 29.15 & 4.94 & -2.58 \\
\hbox{Fe~I} & 4443.194 & 2.860 & -1.043 & 11.70 & 12.17 & 4.89 & -2.63 \\
\hbox{Fe~I} & 4447.717 & 2.220 & -1.342 & 24.50 & 24.17 & 4.93 & -2.59 \\
\hbox{Fe~I} & 4461.653 & 0.090 & -3.210 & 40.90 & 40.28 & 5.03 & -2.49 \\
\hbox{Fe~I} & 4466.552 & 2.830 & -0.600 & 29.40 & 28.79 & 4.91 & -2.61 \\
\hbox{Fe~I} & 4489.739 & 0.120 & -3.966 & 11.70 & 10.87 & 5.02 & -2.50 \\
\hbox{Fe~I} & 4494.563 & 2.200 & -1.136 & 34.40 & 33.48 & 4.91 & -2.61 \\
\hbox{Fe~I} & 4531.148 & 1.490 & -2.155 & 22.30 & 22.08 & 4.95 & -2.57 \\
\hbox{Fe~I} & 4602.941 & 1.490 & -2.209 & 21.60 & 20.36 & 4.96 & -2.57 \\
\hbox{Fe~I} & 4736.773 & 3.210 & -0.752 & 14.00 & 13.90 & 4.99 & -2.53 \\
\hbox{Fe~I} & 4890.755 & 2.880 & -0.394 & 35.90 & 34.51 & 4.83 & -2.69 \\
\hbox{Fe~I} & 4918.994 & 2.870 & -0.342 & 38.80 & 37.93 & 4.83 & -2.69 \\
\hbox{Fe~I} & 4938.814 & 2.880 & -1.077 & 12.10 & 11.39 & 4.87 & -2.65 \\
\hbox{Fe~I} & 4994.130 & 0.920 & -3.080 & 13.10 & 13.47 & 5.00 & -2.52 \\
\hbox{Fe~I} & 5001.863 & 3.880 &  0.010 & 13.30 & 13.22 & 4.85 & -2.68 \\
\hbox{Fe~I} & 5006.119 & 2.830 & -0.638 & 27.50 & 28.05 & 4.88 & -2.64 \\
\hbox{Fe~I} & 5012.068 & 0.860 & -2.642 & 32.80 & 32.77 & 5.03 & -2.49 \\
\hbox{Fe~I} & 5041.072 & 0.960 & -3.087 & 13.20 & 13.81 & 5.06 & -2.46 \\
\hbox{Fe~I} & 5049.820 & 2.280 & -1.355 & 22.60 & 22.32 & 4.92 & -2.60 \\
\hbox{Fe~I} & 5051.635 & 0.920 & -2.795 & 22.80 & 22.86 & 5.01 & -2.51 \\
\hbox{Fe~I} & 5068.766 & 2.940 & -1.042 & 11.20 & 11.68 & 4.91 & -2.61 \\
\hbox{Fe~I} & 5083.338 & 0.960 & -2.958 & 15.30 & 15.96 & 5.01 & -2.51 \\
\hbox{Fe~I} & 5098.698 & 2.180 & -2.026 & 10.20 & 10.55 & 5.08 & -2.44 \\
\hbox{Fe~I} & 5110.413 & 0.000 & -3.760 & 24.10 & 24.79 & 5.09 & -2.43 \\
\hbox{Fe~I} & 5123.720 & 1.010 & -3.068 & 11.00 & 11.48 & 5.00 & -2.52 \\
\hbox{Fe~I} & 5162.273 & 4.180 &  0.020 & 12.60 & 12.18 & 5.08 & -2.44 \\
\hbox{Fe~I} & 5166.282 & 0.000 & -4.195 & 10.30 & 10.58 & 5.06 & -2.46 \\
\hbox{Fe~I} & 5171.596 & 1.490 & -1.793 & 41.20 & 41.63 & 4.99 & -2.53 \\
\hbox{Fe~I} & 5191.454 & 3.040 & -0.551 & 21.90 & 21.78 & 4.84 & -2.68 \\
\hbox{Fe~I} & 5192.344 & 3.000 & -0.421 & 28.40 & 28.72 & 4.84 & -2.68 \\
\hbox{Fe~I} & 5194.941 & 1.560 & -2.090 & 24.40 & 24.25 & 4.97 & -2.55 \\
\hbox{Fe~I} & 5216.274 & 1.610 & -2.150 & 19.60 & 19.75 & 4.96 & -2.56 \\
\hbox{Fe~I} & 5266.555 & 3.000 & -0.386 & 32.90 & 30.80 & 4.84 & -2.68 \\
\hbox{Fe~I} & 5281.790 & 3.040 & -0.834 & 13.50 & 13.37 & 4.85 & -2.67 \\
\hbox{Fe~I} & 5283.621 & 3.240 & -0.432 & 19.50 & 19.20 & 4.84 & -2.68 \\
\hbox{Fe~I} & 5302.300 & 3.280 & -0.720 & 10.20 & 10.05 & 4.84 & -2.68 \\
\hbox{Fe~I} & 5324.179 & 3.210 & -0.103 & 35.40 & 33.05 & 4.81 & -2.71 \\
\hbox{Fe~I} & 5339.929 & 3.270 & -0.647 & 12.70 & 12.52 & 4.86 & -2.67 \\
\hbox{Fe~I} & 5367.466 & 4.410 &  0.443 & 11.30 & 10.98 & 4.82 & -2.70 \\
\hbox{Fe~I} & 5369.961 & 4.370 &  0.536 & 14.10 & 13.86 & 4.81 & -2.72 \\
\hbox{Fe~I} & 5383.369 & 4.310 &  0.645 & 18.40 & 18.68 & 4.80 & -2.72 \\
\hbox{Fe~I} & 5393.167 & 3.240 & -0.715 & 12.20 & 11.82 & 4.87 & -2.65 \\
\hbox{Fe~I} & 5404.151 & 4.430 &  0.523 & 18.40 & 18.41 & 5.03 & -2.49 \\
\hbox{Fe~I} & 5415.199 & 4.390 &  0.642 & 17.50 & 16.14 & 4.79 & -2.73 \\
\hbox{Fe~I} & 5569.618 & 3.420 & -0.486 & 12.90 & 12.50 & 4.83 & -2.69 \\
\hbox{Fe~I} & 5572.842 & 3.400 & -0.275 & 19.70 & 19.06 & 4.82 & -2.70 \\
\hbox{Fe~I} & 5615.644 & 3.330 &  0.050 & 34.10 & 33.91 & 4.79 & -2.74 \\
\hbox{Fe~I} & 6252.555 & 2.400 & -1.687 & 11.30 & 11.43 & 4.96 & -2.56 \\
\hbox{Fe~II} & 4178.862 & 2.580 & -2.500 & 18.50 & 18.03 & 4.99 & -2.53 \\
\hbox{Fe~II} & 4416.830 & 2.780 & -2.410 & 10.90 & 10.98 & 4.82 & -2.70 \\
\hbox{Fe~II} & 4508.288 & 2.860 & -2.250 & 16.20 & 15.66 & 4.91 & -2.61 \\
\hbox{Fe~II} & 4515.339 & 2.840 & -2.450 & 11.80 & 11.91 & 4.96 & -2.56 \\
\hbox{Fe~II} & 4520.224 & 2.810 & -2.600 & 11.20 & 11.17 & 5.04 & -2.48 \\
\hbox{Fe~II} & 4583.837 & 2.810 & -1.860 & 37.40 & 36.22 & 4.99 & -2.53 \\
\hbox{Fe~II} & 5234.625 & 3.220 & -2.279 & 12.90 & 13.16 & 5.17 & -2.35 \\
\end{longtable}
}

\end{appendix}

\end{document}